\documentclass{iau}
\pdfoutput=1
\usepackage{graphicx}
\usepackage{psfig}
\usepackage{epsf}

\title[The Supernova -- ISM/Star-formation Interplay] 
{The Supernova -- ISM/Star-formation interplay}

\author[Gerhard Hensler]  %% give here short author list %%
{Gerhard Hensler
}

\affiliation{Department of Astrophysics, University of Vienna,
Tuerkenschanzstr. 17, 1180 Vienna, Austria \\
email: gerhard.hensler@univie.ac.at}

\pubyear{2013}
\volume{296}  %% insert here IAU Symposium No.
\pagerange{119--126}
\date{?? and in revised form ??}
\jname{Supernova environmental impacts}
\editors{A. Ray \& R.A. McCray, eds.}

\begin{document}
\def\HI{H{\sc i} }
\def\HII{H{\sc ii} }
\def\Ha{{\rm H}\alpha }
\def\SSF{\Sigma_{\rm SFR} }
\def\SHa{\Sigma_{\Ha} } 
\def\Sg{\Sigma_{\rm g} }
\def\Smol{\Sigma_{\rm mol} }
\def\rSF{\rho_{\rm SF} }
\def\rg{\rho_{\rm g} }
\def\Msun{M_{\odot} }
\def\sfr{$\Msun yr^{-1}$}
\def\Mpc2{\Msun/pc^2}
\def\tff{\tau_{\rm ff} }
\def\tSF{\tau_{\rm SF} } 
\def\e{\epsilon} 
\def\eSN{\epsilon_{\rm SN} }
\def\eSF{\epsilon_{\rm SF} }
\def\OH{{12\-+log(O/H)} }
\def\lNO{log(N/O) }
\def\cd{chemo-dynamical }
\def\AA{A{\rm \&}A}

\maketitle

\begin{abstract}
Supernovae are the most energetic stellar events and influence the 
interstellar medium by their gasdynamics and energetics. 
By this, both also affect the star formation positively and negatively.
In this paper, we review the complexity of investigations
aiming at understanding the interchange between supernova explosions
with the star-forming molecular clouds. 
Commencing from analytical studies the paper advances to numerical models
of supernova feedback from superbubble scales to galaxy structure.
We also discuss parametrizations of star-formation and 
supernova-energy transfer efficiencies. Since evolutionary models
from the interstellar medium to galaxies are numerous and are applying
multiple recipes of these parameters, only a representative selection
of studies can be discussed here.
%% add here a maximum of 10 keywords, to be taken form the file <Keywords.txt>
\end{abstract}

\firstsection % if your document starts with a section,
              % remove some space above using this command.

\section{Introduction}

Stars form from the cool gas of the interstellar medium (ISM)
and couple to their environment during their 
lives already by stellar mass and energy release, the latter 
comprising radiation and stellar wind energy. Nevertheless, 
the most vehement effect to the ISM and whole galaxies is  
contributed at their deaths when massive stars 
explode as supernovae type II (SNeII) and intermediate-mass
stars expel planetary nebulae or die as SNeIa from binary systems.
Only a minor fraction of the initial star mass is retained 
as remnants, almost 10-15\% for the massive and 20-30\% for 
intermediate-mass stars (\cite{wei83}) and zero in the SNIa case. 
The rest refuels the ISM. 
This cosmic matter cycle acts on intra-galactic scales and
contributes not only energy but also nucleosynthesis products
to the ISM (e.g. \cite{hen10r}). 
The processes which determine this galactic ecosystem seem to be 
fine-tuned in a manner that e.g. galactic gas disks are mostly in 
energy balance in which the vertically integrated 
star-formation rate (SFR) $\SSF$ (in units of \sfr $pc^{-2}$) 
correlates with the gas surface density $\Sg$ over orders of 
magnitude, known as Kennicutt-Schmidt law (\cite{ken98}). 
More precisely, this relation holds for the molecular gas $\Smol$ 
(\cite{sch11}) what can be understood because the molecular gas 
fraction is determined by the ISM pressure and, by this, also the 
star-formation efficiency (SFE) $\eSF$ (\cite{ler08}).
 
If the SFR is simply determined by the molecular gas reservoir and 
by the free-fall time $\tff$ of molecular clouds (\cite{elm02}), it
would exceed the observed one in the Milky Way by up to two orders
of magnitude (e.g. \cite{hen11}), energetic processes have to 
intervene and to stretch the star-formation (SF) timescale with 
respect to the dynamical timescale implying the SFE such that 
%$\tSF = \eSF^{-1} \cdot \tff$. 
$\delta\rSF/\tSF = \eSF \cdot [\rg/\tff]$. 
This equilibrium on disk scales requires that heating processes 
balance the inherent cooling of the ISM. 
Besides multiple heating processes from dissipation of dynamics, 
as e.g. differential disk rotation, gas infall, tidal interactions, 
etc., the above-mentioned local and immediate feedback by formed 
stars themselves is the most favourable mechanism of SF regulation. 
\cite{koe95} demonstrated already that the SFR achieves 
a dependence on $\rg^2$, if the stellar heating is compensated by 
collisional-excited cooling emission (e.g. \cite{bh89}).

\section{Supernova feedback} 

\subsection{Supernovae and the Matter Cycle with Star Formation}

The most efficient stellar energy power is exerted by SNe,
of which those SNeII accumulate to superbubbles
because of their local concentration still in the star-forming sites 
and their short lifetimes, while SN type Ia occur as isolated 
effects on long timescales (\cite{mat01}) and are more distributed 
over the ISM on larger scales. 

After the confirmation of the existence of a hot ISM phase as 
predicted by \cite{spi56}
and an observational baseline of SN remnants (SNR) over decades
(\cite{wol72}), the importance to understand SNR (\cite{che74}) and 
their relevance for the energy, dynamics, and mass budget of the 
ISM phases (\cite{MO77}) and for the non-dynamical matter cycle 
as interplay of gas phases (see e.g. \cite{hab81,ike83}) moved into 
the focus of ISM and galactic research in the 70's.
With a toy model consisting of 6 ISM components and at least 
10 interchange processes \cite{ike84} included SF from giant 
molecular clouds after their formation from cool clouds, which are 
swept-up and condensed in SN shells. By this, SF and SN explosions 
together with different gas phases form a consistent network
of interaction processes. As a reasonable effect of this
local consideration the SF oscillates with timescales determined 
by the gas density and the interaction strengths.  

Taking single SNR models into account, e.g. \cite{che74,cio91} 
modelled the evolution and volume filling of
randomly distributed and temporally exploding SNe with SNR cooling 
and expansion and with mutual transitions between the
warm and cool gas phases. 
Since it is obvious that this hot SN gas dominates energeticly
and kineticly the ISM, the detailed understanding of its interaction 
with the cool gas, its cumulative effect as superbubbles, and the 
dynamical structuring of galactic gas disks 
(comprehensively reviewed by \cite{spi90}) 
is of vital importance for the evolution of galaxies. Because 
SN interactions happen on largely different length and time scales 
such studies have to cover a large variety of aspects (\cite{che77})
reaching from the stirr-up of the ISM by turbulence (this sect.), 
by this, regulating the galactic SF, to the triggering 
of SF (sect. 2.2), and at least to the gas and element loss from 
galaxies (sect.~3).  

First dynamical approaches to the structure evolution of the ISM and
gas disks, aiming at understanding the disk-halo connection were 
performed by \cite{ros95}. 
As heating sources they took the energy of massive stellar winds 
only into account, but overestimated their impact on the ISM because 
highly resolved numerical studies reveal surprisingly 
low energy transfer efficiencies (\cite{hen07}). 
Nonetheless, their models show the compression of gas filaments 
and the expulsion of gas vertically from the disk even under 
self-gravity.

\begin{figure}[ht]
%\hspace{-4cm}
\vspace{-8cm}
\begin{center}
\includegraphics[width=10cm]{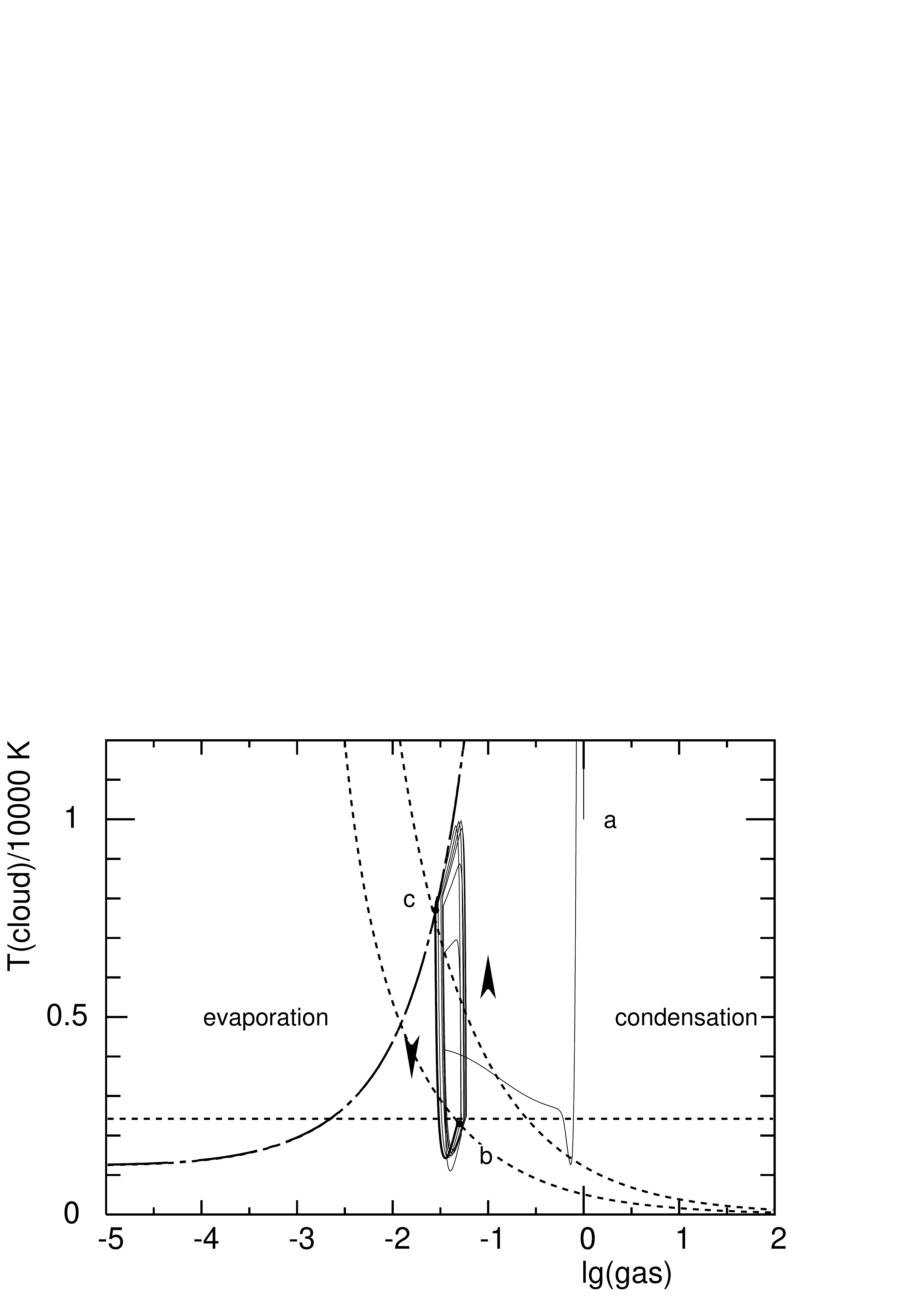} 
\end{center}
%\vspace{6cm}
\caption{Evolution of the full system of cold+hot gas 
from the initial state (`a') until the completion of 
the first few oscillations (solid line). The dashed
lines descending to the right are the loci where the system 
switches from evaporation to condensation (lower curve, at `b') 
and vice versa (at `c'). 
The horizontal dashed line is the locus of the evaporation
funnel and the dot-dashed curve depicts the condensation funnel. 
\it{ (for details see \cite{koe98}).}
}
   \label{fig1}
\end{figure}

The influence of SNe on the SF can be imagined by two 
measures: the expansion of hot gas, its deposit of turbulent energy
(\cite{ml04}), and its evaporation of embedded cool gas should, 
at first, lead to the suppression of SF, 
while vice versa the sweep-up and condensation of surrounding gas 
in the shells of SNRs and superbubbles could trigger SF.
That hot SN gas regulates SF thermally has been demonstrated 
by \cite{koe98} who analysed the equations of a multi-component system 
when thermal conduction accounts for gas-phase transitions 
(see fig. \ref{fig1}).

%---------------------------
%---------------------------

In more comprehensive numerical investigations of the ISM evolution 
\cite{sly05} studied the influence of SF and SN feedback on the 
SFR with self-gravity of gas and stars. 
As the main issues one can summarize that feedback enhances the ISM 
porosity, increases the gas velocity dispersion and the contrasts 
of T and $\rho$, so that smaller and more pronounced structures
form, and, most importantly, that the SFR is by a factor 
of two higher than without feedback. 

At the same time, \cite{avi04} simulated the structure evolution 
of the solar vicinity in a $1 \times 1 \times 10\, kpc^3$ box and 
identified the Local Bubble and its neighbouring Loop I in their 
models as well as the vertical matter cycle.
In addition, filamentary neutral gas structures become visible  
where SF of low-mass star clusters (see next sect.) resulting 
from the production of gas shells by local SNe express the positive
SN feedback. Since their ISM processes do not include SF 
self-regulation processes by stellar radiation and
winds as well as heat conduction, they lack of negative SF feedback.

\subsection{Star-formation triggering}

SN and stellar wind-driven bubbles sweep up surrounding 
gas, condense it, and could, by this, trigger SF in a 
self-propagating manner as a positive feedback. 
The perception of SF trigger in SN or superbubble shells sounds
reasonable from the point of view of numerical models because 
shock front compression as shown by \cite{che74} and sufficient 
swept-up mass from the ambient ISM (as shown in fig. 2
preferably in the gas disk itself) lead to cooling and 
gravitational instabilities. This mechanism, however, is not
generally confirmed by observations. 
Shell-like distributions of young stars, are e.g. found in
G54.4-0.3, called {\it sharky} (\cite{jun92}), 
in the Orion-Monoceros region (\cite{wil05}), 
more promising in the Orion-Eridanus shell (\cite{lee09}),
and in several superbubbles in the Large Magellanic Cloud,
as e.g. Henize 206 (\cite{gor04}). 
The SF associated with the SNR IC443 (\cite{xu11}) raises e.g. 
the already above-mentioned question, whether SNR-triggered
SF is capable to lead to massive star clusters which fill the
whole stellar mass function equally, because here only about 
10$^4 \Msun$ of molecular gas is involved. 
Also the formation of Gould's Belt as site of low-mass stellar 
associations in the shell of a superbubble is most probable 
(\cite{mor99}).

%---------------------------
\begin{figure}[ht]
%\begin{center}
\begin{tabular}{lr}
\resizebox{0.6\columnwidth}{!}{%
\includegraphics[width=5cm]{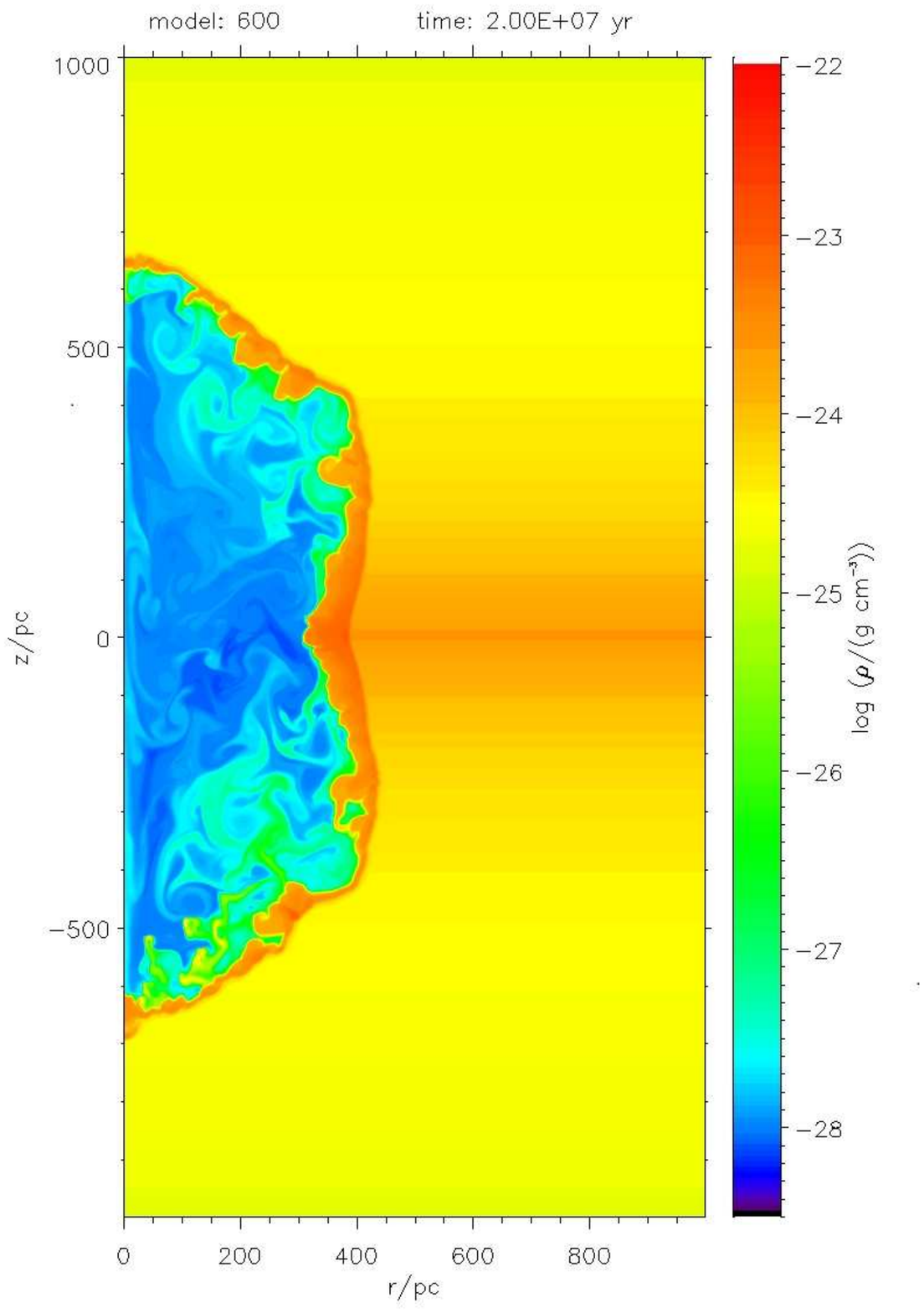} }
%\end{center}
&
\begin{minipage}[t]{5.cm}
\vspace{-10.5cm}
\caption{
Density distribution of a superbubble after almost 20 Myrs.
The superbubble results from 100 supernova typeII explosions of 
stars in the mass range between 10 and 100 $\Msun$. The temporal 
sequence of explosions happens according to the lifetimes of massive
stars with a Salpeter IMF. The star cluster is located at the origin 
of the coordinates. The galactic disk is vertically composed of the
three-phase interstellar medium, cool, warm, and hot phase,
respectively, with 
(central density $\rho_0\, [in\, g\, cm^{-3}]$; 
temperature T [{\it in K}]; 
scaleheight H [{\it in pc}]) of 
($2\times 10^{-24}$; 150; 100), 
($5\times 10^{-25}$; 9000; 1000)
($1.7\times 10^{-27}$; $2\times 10^6$; 4000).
The density varies from almost $10^{-23}\, g\, cm^{-3}$, in the 
densest part of the shell to $2\times 10^{-28}\, g\, cm^{-3}$ in
the darkest bubble interiors. {\it (from \cite{gud02})}
}
\end{minipage}
\hfill
\end{tabular}
\vspace{-1.0cm}
   \label{fig2}
\end{figure}
%---------------------------

Such SF trigger by the condensation of swept-up gas in SN or more
efficiently in superbubbles (see fig. 2) can be explored in detail 
by the investigation of the fragmentation timescale 
(see e.g. \cite{ehl97,fuk00}). Ehlerova et al. 
compared a self-similar analytical solution with the results of 3D 
numerical simulations of superbubble expansions in homogeneous media. 
The amount of energy supply from the final number of young stars in an 
OB association, the value of the sound speed, the stratification and 
density of the ambient medium, the galactic differential rotation,
and the vertical gravitational force in the galactic disk, all these 
influence the fragmentation. The typical superbubble radius, at 
which shells start to fragment, decreases from almost 700 pc at an
ambient gas density of 1 cm$^{-3}$ to 200 pc at 10 cm$^{-3}$. 
While in thick disks like they exist in DGs nearly the whole shell 
fragments, so that the SF may propagate in all directions,  
in thin disks it is restricted to gas layer around the 
galactic equator only. 
Since the applied thin shell approximation is reasonably only a 
0th-order approach, in a recent paper \cite{wue10} 
clarify that the shell thickness and the environmental pressure 
influences the fragments in the sense that their sizes become smaller 
for higher pressure. Nevertheless, the deviations from the thin-shell 
approximations are not large.

Since these studies do not allow for the inhomogeneity of the ISM,
another possible feedback effect by SN can be caused, when the 
ultra-fast SNR shock overruns a dense interstellar cloud,
so that the clouds are quenched (\cite{orl05}). Stars should be 
formed instantaneously by such cloud crushing, and the cloud
mass determines the star cluster mass.

\subsection{Supernova energy impact}

Although numerical simulations have been performed to understand 
the heating (or energy transfer) efficiency $\eSN$ of SNe (\cite{tho98}), 
superbubbles (e.g. \cite{str04}), and starbursts (\cite{mel04}), they 
are yet too simplistic and mostly spatially poorly resolved to account
for quantitative results. 
Thornton et al. derived an efficiency $\eSN$ of 0.1 from 1D 
SN simulations as already applied by chemo-dynamical galaxy models 
(\cite{sam97}), while unity is also used in some galaxy models 
(see sect. 3), but seems far too large.

\section{Supernovae and Galaxy Evolution}

\subsection{Supernova parametrization in numerical models}

Because of their high power in various forms, massive stars are 
usually taken into account as the only heating sources for the ISM 
in galaxy evolution models and here mostly only the energy deposit 
of SNII explosions alone. Recently, \cite{sti13} envoked indeed the 
necessity of short-term SF feedback which is naturely implied by
massive stellar radiation and winds but was already included in the
chemo-dynamical prescription (see e.g. \cite{sam97}).
It must be explored, however, how effectively this energy is 
transferred into the ISM as turbulent and finally as thermal energy. 
Although it is generally agreed that the explosive energy $E_{SN}$
of an individual SN lies around 10$^{51}$ ergs with significant 
uncertainties of probably one order of magnitude, 
the energy deposit is still more than unclear, but is one of the 
most important ingredients for galaxy formation and evolution
(e.g. \cite{efs00,sil03}). Massive stars do not disperse from their 
SF site and thus explode within the stellar associations, by this,
contributing significantly to the ISM structure formation 
e.g. by cavities and holes in the \HI gas and chimneys of hot gas. 
On large scales SNeII trigger the matter cycle via galactic outflows 
from a gaseous disk.
By this, also the chemical evolution is affected thru the loss of 
metal-enriched gas from a galaxy (for observations see e.g. 
\cite{mar02}, for models e.g. \cite{rec06b}). 

Although numerical experiments of superbubbles and galactic winds 
are performed, yet they only demonstrate the destructive effect 
on the surrounding ISM but lack of self-consistency and a complex treatment. 
Simulations of the chemical evolution of starburst DGs by 
\cite{rec02}, 2006a), that are denoted to reproduce the peculiar 
abundance patterns in these galaxies by different SF episodes, 
found, that $\eSN$ can vary widely: it starts with 10\% drops and 
increases successively but not above 18\% (\cite{hen11}). 
For the subsequent SNIa explosions, always single events, $\eSN$ is 
much smaller, i.e. below 1\%.
Moreover, if a closely following SF episode (might be another burst) 
drives its SNII explosions into an already existing chimney of a 
predecessor superbubble, the hot gas can more easily escape 
without any hindrance and thus affects the ISM energy budget much less.
\cite{rec06a} found that depending on the galactic \HI density the 
chimneys do not close before a few hundred Myrs. 

Since $\eSF$ must inherently depend on the 
local conditions so that it is high in bursting SF modes, 
but of percentage level in the self-regulated SF mode, 
numerical simulations often try to derive the 
''realistic'' SFE by comparing models of largely different 
$\eSF$ with observations, as e.g. to reproduce gas structures in 
galaxy disks and galactic winds (e.g. undertaken by 
\cite{tas06}, 2008) with $\eSF$ = 0.05 and 0.5). 
In addition, specific SN energy deposit $\eSN \times E_{SN}/M_*$ 
by them is fixed to 10$^{51}$ erg per 55 $\Msun$ of formed stars  
($1.8\times 10^{49}$ erg $\Msun^{-1}$ by \cite{dvs08}), but their
results cannot ye tbe treated quantitatively, since they also mismatch 
with the Kennicutt relation.
\cite{sti06} performed a comprehensive study about the
influence of $\eSN$ over a large scale of values and also of the
dependence on the mass resolution of their SPH scheme GASOLINE
of galaxy evolution. Although their results demonstrate the already 
expected trends, more realistic treatments must adapt $\eSN$ 
self-consistently to the local state of the ISM what requires
much larger numerical capacity for more intensive high-resolution 
numerical simulations. 

Theoretical studies by \cite{elm97} achieved a dependence of $\eSF$ 
on the external pressure, while \cite{koe95} explored a temperature 
dependence of the SFR both effects affecting the SFR. 
Furthermore, most galaxy evolutionary models at present lack 
of the appropriate representation of the different ISM phases 
allowing for their dynamics and their direct interactions by 
heat conduction, dynamical drag, and dynamical instabilities 
thru forming interfaces, not to mention resolving the turbulence
cascade.

\subsection{Superbubbles and galactic winds}

%---------------------------
\begin{figure}[ht]
\begin{center}
\includegraphics[width=9cm]{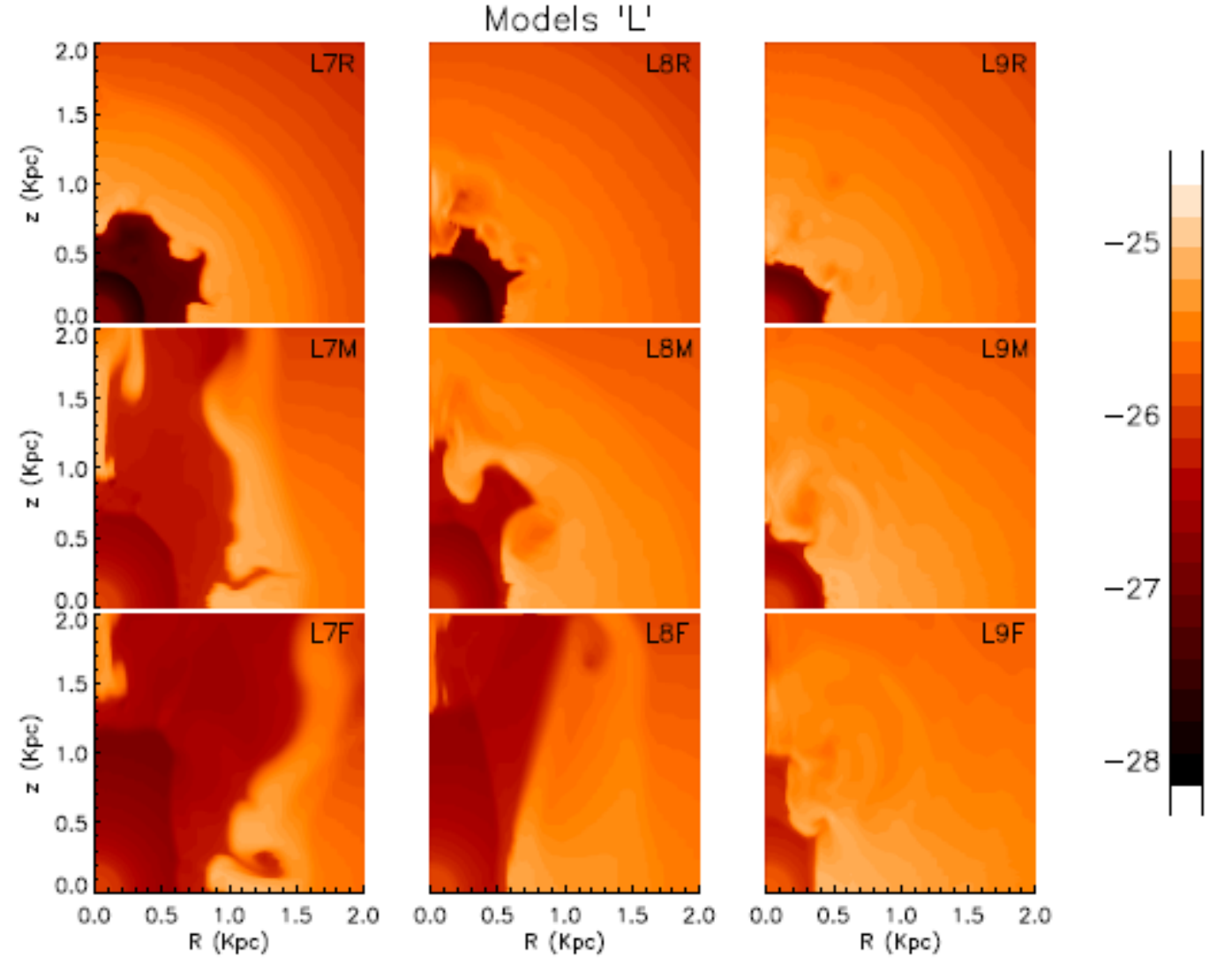} 
\end{center}
\caption{Gas density distribution for the 9 models of the of 
60\%-initial gas fraction ("L") after 200 Myr of evolution. 
The first column represents models with 10$^7 \Msun$ of initial 
baryonic mass; the middle column shows the gas distribution of
10$^8 \Msun$ models, and the right column the 10$^9 \Msun$ models. 
The top row models are characterized by a roundish initial gas 
distribution ("R"; with b/a = 5), the middle row by b/a = 1 ("M"), 
and finally the bottom row represents flat disks with 
b/a = 0.2 ("F"). 
At the top-right corner of each panel the model designation is also 
indicated. The right-hand strip shows the (logarithmic) density scale 
(in g cm$^{-3}$). \it{(from \cite{rec13}) }
}
   \label{wind}
\end{figure}
%---------------------------

A superbubble expanding from a stellar association embedded in a 
\HI disk has, at first, to act against the surrounding medium, 
by this, is cooling due to its pressure work and radiation, 
but compresses the swept-up shell material and implies turbulent 
energy to the ISM. 
How much the superbubble expansion is efficiently hampered depends 
on the surrounding gas density and pressure, the \HI disk shape 
(\cite{rec13}), and the energy loss by radiative cooling. 
From fig.\ref{wind} it is discernible
that only flat gas disks of preferably low-mass galaxies allow a 
hot wind to escape from the galaxy. Consequences for the chemical 
evolution are in the focus of those models (\cite{rec13}).
And finally, observed superbubbles also reveal a mismatch of their
spatial extent and the energy content required to drive the 
expansion with the observed X-ray luminosity (\cite{hen98}). 
This fact can be only explained by a significant energy loss.

\section{Conclusions}

The dominating influence of SN explosions on structure,
dynamics, and energy budget of the ISM are obvious and agreed.
Signs and strengths of these feedback effects are, however, widely 
uncertain. Whether the feedback is positive (trigger) or negative 
(suppression) can be understood analyticly from first principles,
but because of the non-linearity and the complexity level of the acting
physical plasma processes clear results cannot be quantified 
reliably. In addition, the temporal behaviour varies by orders of 
magnitude because of the changing conditions.
In summary, the energy transfer efficiency of SN energy to the ISM 
energy modes is much below unity and must not be overestimated, 
but also depends on the temporal and local conditions.

\begin{acknowledgments}
The author is grateful for numerous discussions with Simone Recchi 
and with the participants of the Aspen Center for Physics summer program
2012 on "Star-formation regulation" which was supported in part
by the NSF under Grant No. PHYS-1066293. 
\end{acknowledgments}

\begin{discussion}
\discuss{Cesarsky}{Cosmic Ray and magnetic fields potentially 
play a big role in 
some of the process you were discussing, fragmentation, bubble formation and especially galactic winds. 
Were they considered in some of the models you discussed?}
\discuss{Hensler}{Not yet. The actually treated complexity of 
the ISM processes of multi phases, SF, and star-gas interactions
is already challenging the high-performance computing capacity.}
\discuss{Zhou, P.}{Is there any clear observational evidence to prove that star formation could be triggered by SNR? How to distinguish the SNR impacts from progenitor wind impact?}
\discuss{Hensler}{This observational evidence seems to exist, 
but obviously not in general. This means that local conditions
determine the possibility of SF triggering. From models of
wind-blown and radiation-driven \HII regions the 
stellar wind impact seems not to sweep-up and compress the
surrounding gas sufficiently; see e.g. Freyer, Hensler, \& Yorke, 2003, \textit{ApJ}, 594, 888 and 2006, \textit{ApJ}, 638, 262.
}
\end{discussion}
\end{document}